\documentclass[twocolumn,showpacs,superscriptaddress,twoside,pra]{revtex4}
\usepackage{amsmath,amssymb,url,graphicx,epstopdf,bm}
\usepackage{epsfig,hyperref,amsmath,amssymb,amsfonts, latexsym, dsfont, wasysym,multirow,mathrsfs,xcolor,color}
\usepackage[caption=false]{subfig}
\usepackage{pifont}

\begin{document}
\title{Evolutionary algorithms for hard quantum control}
\author{Ehsan Zahedinejad}
\affiliation{%
	Institute for Quantum Science and Technology,
	University of Calgary, Alberta, Canada T2N 1N4}
\author{Sophie Schirmer}
\affiliation{%
	College of Science, Swansea University, Singleton Park,
	Swansea SA2 8PP, Wales, United Kingdom}
\affiliation{%
	Kavli Institute for Theoretical Physics,
	University of California at Santa Barbara, California 93106-4030, USA}
\author{Barry C. Sanders}\email{sandersb@ucalgary.ca}
\affiliation{%
	Institute for Quantum Science and Technology,
	University of Calgary, Alberta, Canada T2N 1N4}
\affiliation{%
	Kavli Institute for Theoretical Physics,
	University of California at Santa Barbara, California 93106-4030, USA}
\affiliation{%
	Program in Quantum Information Science, Canadian Institute for Advanced Research,
	Toronto, Ontario M5G 1Z8, Canada}
\affiliation{%
	Hefei National Laboratory for Physical Sciences at the Microscale
		and Department of Modern Physics,
	University of Science and Technology of China, Anhui 230026, China}
\begin{abstract}
Although quantum control typically relies on greedy (local)
optimization, traps (irregular critical points) in the control
landscape can make optimization hard by foiling local search
strategies.  We demonstrate the failure of greedy algorithms as well
as the (non greedy) genetic-algorithm method to realize two fast
quantum computing gates: a qutrit phase gate and a controlled-NOT
gate.  We show that our evolutionary algorithm circumvents the trap to
deliver effective quantum control in both instances.  Even when greedy
algorithms succeed, our evolutionary algorithm can deliver a superior
control procedure, for example, reducing the need for high time
resolution.
\end{abstract}

\date{\today}
\maketitle

%
%

\section{Introduction}
\label{sec:introduction}
Quantum control aims to steer quantum dynamics towards closely realizing
specific quantum states or operations~\cite{SB03,DP10} with applications
to femtosecond lasers~\cite{ABB+98,MS98}, nuclear magnetic
resonance~\cite{NMR1,KRK+05} and other
resonators~\cite{RSK05,MVT98,HJH+03}, laser-driven molecular
reactions~\cite{BS92,TR85}, and quantum gate synthesis for quantum
computing~\cite{SSK+05}.  Control is achieved by varying the strengths
of different contributing processes (external fields) over time such
that the resultant evolution closely approximates the desired evolution.
The quality of a given quantum control procedure is quantified by its
fitness~\cite{dFS13} such as fidelity or distance for the
approximated quantum state~\cite{EW14} or quantum
gate~\cite{MMG+10} and the target state or gate.

A key goal in quantum control is to reach the fittest procedure
possible within the target time~$T$ subject to certain resource limits
such as limiting the number of independent control parameters~$K$ and
therefore the time resolution~$T/K$ for time-domain quantum control.
Practical considerations usually tightly constrain the maximum
allowable values for~$T$, and lower bounds for~$T$ are central to
questions about fundamental ``quantum speed limits'' to operations in
quantum computing, quantum metrology, and quantum
chemistry~\cite{LT09,TEDdM13,dCEPH13,Heg13}

Choosing control parameters to maximize the procedure fitness is an
optimization problem.  Early quantum control employed non greedy
approaches, e.g., the genetic algorithm (GA)~\cite{BYW+97,Gol89}.
Today greedy algorithms dominate the methodology as local optimization
strategies usually have lower computational cost than global search
algorithms and the fitness landscape (plot of fitness vs.\ control
parameters) typically appears to be tractable~\cite{RHR04}.
Unfortunately, greedy algorithms can fail even for low-dimensional
quantum control with simple Hamiltonians if~$T$ must be short.  This
seemingly innocuous constraint eliminates any guarantee of global
optimality for local extrema.

Although it is tempting to attribute failure to find a satisfactory
control procedure to infeasibility of the constrained problem, we show
that this failure can instead be due to restricting strategies to
greedy algorithms.  To make our case, we present examples of control
problems involving simple systems for which greedy algorithms
overwhelmingly fail. These two control problems are
especially contrived to be hard to solve using common quantum-control
techniques, but the problems are physically meaningful as discussed
  in Secs.~\ref{subsec:qutrit} and~\ref{subsec:CNot}, respectively.
  We use the term ``hard'' to refer to problems that defy existing methods
  in the sense that the probability that they produce a satisfactory
  solution is small. 
  A key element of these problems is that the time required for the 
  unitary operation is short, which could make many quantum control problems
  hard. We show that these hard quantum control problems can be
solved using global optimization techniques based on the
differential evolution (DE) algorithm~\cite{SP97}, which succeeds in
finding effective controls up to the computational-power limits
(machine error) even for very short~$T$ and very few controls.

We compare greedy vs non greedy algorithms for realizing two different
quantum computing gates: the original qutrit phase
gate~\cite{GKP01,BdGS02} and the two-qubit controlled-Not
(\textsc{CNOT}) gate~\cite{BBC+95}, which are key elements of standard
quantum computing instructions sets for qutrits and for qubits,
respectively.  We show that, for each gate and given our selected
drift and control Hamiltonians, the greedy algorithm fails to find a
high-fitness quantum-control procedure for short target time~$T$ while
our non greedy DE algorithm succeeds.  Moreover, for larger~$T$ where
greedy algorithms work, DE is able to find solutions requiring fewer
independent control parameters~$K$ than the greedy algorithms tested.
Interestingly, the common non greedy GA also strongly fails for our
test problems.

\section{Quantum control}
\label{sec:QCS}
In any quantum-control problem, the goal is to decompose the
  system's Hamiltonian into a controllable and an uncontrollable part
  and steering the dynamics towards a desired evolution through
  varying the controllable part of the system.  Here we first explain
  the system Hamiltonian in the context of control theory and then
  discuss our choice of the fitness function serving as the objective
  function for the purpose of optimization.

\subsection{Quantum control Hamiltonian}
\label{subsec:quantumcontrolH}
For a closed system, the Hamiltonian
\begin{equation}
\label{eq:controlHamiltonian}
	\hat{H}\left[\varepsilon(t)\right]
		=\hat{H}^\text{dr}+\bm{\varepsilon}(t)\cdot\bm{\hat{H}}^\text{c}
		=\hat{H}^\text{dr}+\sum_{\ell=1}^L\varepsilon_\ell(t)\hat{H}^\text{c}_\ell,
\end{equation}
acts on Hilbert space~$\mathscr{H}$~\cite{dAl07}
with drift Hamiltonian~$\hat{H}^\text{dr}$
describing free (uncontrolled) evolution, which we
treat as being time-independent here.
The control Hamiltonians,
represented by the vector operator $\bm{\hat
H^\text{c}}(t)=(\hat{H}^\text{c}_\ell)$ (for~$\{\ell\}$ the control
field labels), should steer the system towards the desired evolution with
time-varying (here piecewise constant) control amplitudes
contained in the vector~$\bm{\varepsilon}(t):=\{\varepsilon_\ell(t)\}$.

The resultant unitary evolution operator is formally
\begin{equation}
\label{eq:U}
	U\left[\bm{\varepsilon(t)};T\right]
		=\mathcal{T}\exp\left\{-\text{i}\int_0^T\hat{H}(\bm{\varepsilon}(t))\text{d}t\right\}
\end{equation}
with~$\mathcal{T}$ the time-ordering operator~\cite{DGT86}.
We aim to approximate a target unitary evolution operator~$U$ within
duration~$T$ by a unitary operator
$\tilde{U}[\bm{\varepsilon(t)};T]$ with minimum distance
\begin{equation}
\label{eq:distance}
	\|U-\tilde{U}[\bm{\varepsilon}(t);T]\|
\end{equation}
between the realized evolution and the target.

\subsection{Fitness functional}
\label{subsec:FF}

The quality of a candidate quantum control procedure is quantified by
the fitness functional
\begin{equation}
\label{eq:F}
	\mathscr{F}[\bm{\varepsilon}(t)]
		=\mathscr{F}(U[\bm{\varepsilon}(t);T])
		=1-\|U-\tilde{U}[\bm{\varepsilon}(t);T]\|
\end{equation}
with~$\|\bullet\|$ the operator norm and the final term in~(\ref{eq:F})
the trace distance~\cite{GVC12} between the target and the actual evolution
operators.
The optimization problem is to maximize
$\mathscr{F}[\bm{\varepsilon}(t)]$, i.e., to reduce the distance~(\ref{eq:distance}).
For numerical simulation we use the explicit form
\begin{equation}
\label{eq:Objefun}
	\mathscr{F}(t)
		=\frac{1}{N}\text{Re}\left(\text{Tr}
			\left(U[\bm{\varepsilon}(t),T]\tilde{U}[\bm{\varepsilon}(t);T]\right)\right)
\end{equation}
of the fidelity function~\cite{SSK+05} between the target $U[\bm{\varepsilon}(t);T]$ and the approximate unitary operator $\tilde{U}[\bm{\varepsilon}(t);T]$,
with~$N$ the Hilbert-space dimension.
\section{Criteria to Evaluate Algorithm Performance}
\label{sec:PCF}
Evaluating and comparing algorithms for optimization should be conducted
fairly and clearly.  Using run-time directly as a cost criterion
obscures fundamental issues in comparing the intrinsic differences.
Therefore, we
evaluate and compare algorithms based on whether the algorithm yields a
sufficiently optimal solution over many attempts, here called ``runs.''
Each run is allowed to iterate until it succeeds or fails in
which case the run aborts.

The iteration number of run~$r$ is~$\imath$,
and the total number of iterations for run~$r$ is denoted~$I_r$,
with~$I_R$ the maximum iteration number over all~$R$ runs.  For~$R$
runs, we determine and tabulate the best and worst fitness values
obtained over these runs, and we characterize the statistics of error
values according to the median error and the probability~$\wp$, or
percentage, of runs whose error is less than some threshold value.

We compare the performance of the optimization algorithms based on the
bounds and statistics of the statistics of runs, and these statistics
are analyzed in plots that depict the fidelity vs~$\imath$ for each of
the many runs.  A plot of~$\mathscr F$ vs~$\imath$ is 
overly crowded to reveal key features clearly.
Therefore, we stretch the plots by presenting the monotonically related ``logarithmic
infidelity''
\begin{equation}
\label{eq:loginfidelity}
	L:=\log_{10}(1-\text{Re}\left(\mathscr{F})\right)
\end{equation}
vs.~$\imath$ for each run.
Logarithmic infidelity is zero for perfect infidelity
\begin{equation}
\label{eq:perfectinfidelity}
	\mathscr{F}=0,
\end{equation}
hence approximately bounded by~$L=-16$ for double precision
and ideally $-\infty$ for perfect fidelity ($\mathscr{F}=1$).

The algorithm for run~$r$ is deemed successful if the final~$\mathscr{F}_r$
exceeds a minimum threshold~$L^\text{t}$,
which we set to $L_\text{t}=-4$ commensurate with the widely accepted gate fidelity required for
scalable quantum computing~\cite{Ste03}.  
Our algorithm aborts a run after $I_r$ iterations only if
the change in~$\mathscr{F}_r$ is within machine error or an infidelity within machine precision is reached.
The percentage of runs that beat~$L^\text{t}$ is denoted~$\wp^\text{t}$.

  A fair comparison of greedy vs evolutionary algorithms would
  consider an equal number of trials in each case.  In order to make a
  stronger case that our evolutionary algorithm is superior, we are
  giving the greedy algorithms an advantage by allowing them
  twice as many runs as for the evolutionary case.  This allowance is
  feasible because greedy algorithms typically run much quicker, so
  allotting additional time to double the number of greedy runs is not
  onerous in terms of computational time.  Specifically, we run the
  greedy algorithms over~80 trials and the evolutionary algorithms
  over~40 trials. Our choice of~80 and~40 trials, respectively, comes
  from our experience in testing these different algorithms, and these
  numbers correspond to balancing achieving sufficient success
  probability against excessive computational cost.

\section{Methods}
\label{sec:methods}
We begin this section with explaining the choice of control
  function that we use for the purpose of optimizing the external
  field. We then discuss the 
  details of how we use the external field to numerically approximate
the unitary operation in (\ref{eq:U}). The last part of this section
discusses the optimization routines that we have tested on two quantum
control cases. We discuss two classes of optimization routines, namely,
local (greedy) and global (evolutionary) algorithms. Our focus is on
the evolutionary algorithms; we provide a detailed
explanation of each algorithm in the appendix section.

\subsection{Type of control function and control parameters}
Now we discuss how the computation works. Numerically, the fitness
functional~(\ref{eq:Objefun}) is evaluated by discretizing the control function
vector~$\bm{\varepsilon}$ by expressing it as a sum of~$K$ orthonormal
functions over the time domain~$[0,T]$ as:
\begin{equation}
	\bm{\varepsilon}
		:=
		\begin{pmatrix}
			\varepsilon_1 \\ \varepsilon_2\\ \vdots \\ \varepsilon_K 
		\end{pmatrix}
		,
\end{equation}
such that each vector element~$\varepsilon_l$ is constant over
sequential time steps of equal duration $\Delta t=T/\text{K}$.

The~$K$ control parameters refer to choosing various weightings of
these control functions.  For our analysis, these~$K$ orthonormal
functions are non overlapping rectangular functions with identical
durations~$T/K$~\cite{Rao83}; i.e., the control functions are
expressed as a weighted series of time bins.
Each control element is randomly generated from the interval $[-1,1]$ and evolves
through the optimization process toward its best optimal value.

This time-bin discretization is commonly used and justified by the
fact that control pulses on experimental hardware are often limited to
this form, although there are alternatives such as decomposition into
$K$ monochromatic functions to be solved in the frequency domain.  The
rectangular time bins also have computational advantages in that the
time-ordered integral~(\ref{eq:U}) is straightforward to evaluate.

\subsection{Optimizing the control function}
  For any optimization problem, greedy algorithms are the primary
  choice if they can provide a satisfactory result. 
Greedy algorithms locally explore the landscape so
their convergence rate toward a local optimum is much faster than for global optimization algorithms.
If the landscape includes many traps,
most trials striving to find global
  optima become ensconced within local traps, 
in which case greedy algorithms fail
  to find the best solution.
Evolutionary algorithms are specifically designed for non convex optimization problems,
which typically arise when the landscape includes many traps and a global optimum is the goal.
Here we find that greedy algorithms are preferred when there is enough time to
  approximate the unitary operation and enough control parameters to
  search the landscape.
Otherwise, evolutionary algorithms are the
  choice to solve the problems. We discuss these two classes of
  optimization algorithms here.

Greedy algorithms include the Nelder-Mead technique~\cite{ON75}, 
Krotov~\cite{Kro95,ST02,MSG+11},
the quasi-Newton method, which employs the
Broyden-Fletcher-Goldfarb-Shanno (BFGS) approximation of the
Hessian~\cite{Bro70,Fle70,Fle13,Gol70,Sha70}. 
The Nelder-Mead technique uses the fitness functional only, 
whereas the Krotov algorithm uses both the fitness functional and
its gradient~($\boldsymbol\nabla\mathscr{F}$) with respect to control elements,
and quasi-Newton uses
the fitness functional as well as its gradient~($\boldsymbol\nabla\mathscr{F}$)
and Hessian~($\nabla^2\mathscr{F}$)
to find local optima over many iterations.
Lie group techniques can
help to determine the gradient analytically~\cite{SdF11,FdFS12}, and
numerical techniques work generally.

Greedy algorithms are especially successful if both~$T$ and~$K$ are
sufficiently large.  For constrained~$T$, second-order traps such that
the Hessian is negative semidefinite arise~\cite{PT11,RHL+12,PT12note}
and numerical evidence arguably exists for the presence of other
traps~\cite{PT12IJC,dFS13,PT12}.

As an alternative to greedy algorithms, we 
consider evolutionary algorithms for quantum control. Evolutionary
algorithms are stochastic optimization algorithms, inspired by the
process whereby biological organisms adapt and survive~\cite{DK06}.
These algorithms only require the fitness functional and not its
gradient or Hessian.

The large class of evolutionary algorithms includes
simulated annealing~\cite{KGV83},
ant-colony systems~\cite{DBS06},
memetic algorithms~\cite{Mos89}, DE~\cite{SP97},
particle swarm optimization (PSO)~\cite{Ken10}, and GA~\cite{Gol89}~{\it inter alia}~\cite{DK06},
but we choose to test just the three most common or promising evolutionary algorithms,
namely traditional GA (a commonly used algorithm)
and the modern PSO and DE algorithms (promising for this type of problem).
The promising nature of PSO and DE is based on
many studies~\cite{PBC+09,EHG05,EZG+09} that have shown the superiority of DE and PSO
over other evolutionary algorithms.
All PSO, DE, and GA employ the initial condition of multiple guesses,
called particles in PSO and chromosomes in DE and GA. Each test
function evolves iteratively along trajectories in parameter space and
experience different fitness values.

In GA, ``parent'' chromosomes go through three steps~---~selection,
crossover, and mutation~---~to generate a new generation (``offspring'') of
chromosomes.  We tested all MATLAB (version R12118) GA
options and found that the wheel-roulette, two-point and uniform
methods perform best.  We test GA fairly by optimizing population
number~$N$ independently for each GA variant and fix the run time to
equal those for PSO and DE runs. 

In PSO the particle evolves according to a Langevin equation that
includes a random kick, an attractive force to its previous best
fitness, and a force pulling to the particle to the fittest particle in
its neighborhood (where the size of the neighborhood is logarithmic in
the number of particles).  Neighborhoods overlap such that they do not
partition into distinct sets.  Specifically, we employ three PSO
variants labeled here as PSO1, PSO2~\cite{Tre03}, and PSO3~\cite{CK02}.

In DE each chromosome breeds with three other randomly chosen
chromosomes from the same generation to produce a ``daughter,'' and
the fittest of the original vs the bred daughter survives to the
next generation.  We use a DE variant that incorporates mutation
scaling factor~$\mu\in[0,2]$ and cross over rate~$\xi$~\cite{SP97}.
In each generation the difference between two randomly chosen target
vectors is weighted by~$\mu$, then added to a third randomly selected
target vector to generate the new set of vectors called donors.  This
quantity~$\mu$ determines the DE step size for exploring the control
landscape.  Donor vector elements are incorporated into target vectors
with probability~$\xi$ to generate trial vectors, and the fittest of
the target and trial vectors survive to the next generation.

Details of DE, PSO, and GA and a comparison between these three
algorithms can be found in the appendixes.

\subsection{Evaluating the objective function}
\label{subsec:evaluating}

We use the following decomposition approach to construct the gate,
\begin{equation}
\label{eq:U1}
	U[\bm{\varepsilon(t)};T]
		=U_K U_{K-1}U_{K-2}\ldots U_{3}U_{2}U_{1},
\end{equation}
with $U_K=\exp(iH(\varepsilon_l)\Delta t)$ and~$T$ the fixed target
time for the unitary operation.  The next step is to
optimize~$\mathscr{F}$ over~$\bm{\varepsilon}(t)$ within target
time~$T$, keeping the number~$K$ of time bins small.  

\section{Two quantum-control cases}
\label{sec:twocases}

Now we proceed to the two quantum-control cases of a qutrit
  phase gate and a \textsc{CNot} gate.
For each individual problem we
  first test conventional greedy algorithms.
For sufficiently
  large~$T$ for effecting the unitary operation, we show that greedy
  algorithms can rapidly converge to a local optimum, but not
  necessarily a global optimum, which we characterize here as a local
  optimum that meets the infidelity condition $L:=-4$.
We also show that
  reducing the time and the number of control parameters transforms
  the problem to a hard optimization problem for which we employ
  evolutionary algorithms as an alternative approach.
  
\subsection{Qutrit phase gate}
\label{subsec:qutrit}
For a qutrit phase gate the Hilbert space
is~$\mathscr{H}=\text{span}\{|0\rangle,|1\rangle,|2\rangle\}$, and the
target gate is
\begin{equation}
\label{eq:qtarget}
	U =\text{e}^{-\text{i}T\hat{H}^\text{dr}}(\text{e}^{\text{-i}\phi}|0\rangle\langle0|
          -\text{i}\text{e}^{\text{-i}\gamma}|1\rangle\langle1|
          -\text{i}\text{e}^{\text{i}\gamma}|2\rangle\langle 2|) 
\end{equation}
with objective parameters corresponding to the phases~$\phi$
and~$\gamma$ and $\hat{H}^\text{dr}$ is the drift Hamiltonian defined
in (\ref{eq:qutritcontrol}).  As our interest is in hard quantum
control problems, we choose a challenging $T$-dependent drift
Hamiltonian and a single control given by~\cite{dFS13}
\begin{equation}
\label{eq:qutritcontrol}
	\hat{H}^\text{dr}
		=\begin{pmatrix} 1+\frac{\pi}{T}&0& 0\\0&1&0\\0&0&2\end{pmatrix},
	\hat{H}^\text{c}
		=\begin{pmatrix} a&1&0\\1&b&1\\0&1&c\end{pmatrix},
\end{equation}
respectively.

This choice of control and drift Hamiltonian~(\ref{eq:qutritcontrol})
provides a rich lode for studying hard quantum control because, for
any target time~$T$, many choices of~$a$, $b$, $c$, $\phi$,
and~$\gamma$ lead to $\varepsilon(t)\equiv 0$ being a critical point.
The resultant criticality results in
$\text{Re}\{[\mathscr{F}[\bm{\varepsilon}(t)]\}<1$ for which the Hessian
  becomes strictly negative definite.  We consider the specific choice
\begin{equation}
	a=2,\, b=2,\, c=1.
\end{equation}
Then the phase choices
\begin{equation}
	\gamma=\frac{5\pi}{3},\;\sin\phi=-\frac{b+c}{a}\cos\gamma,
\end{equation}
ensure that
\begin{equation}
	\varepsilon(t)\equiv 0
\end{equation}
is a critical point, i.e., a point in which
\begin{equation}
	\boldsymbol\nabla\mathscr{F}[\bm{\varepsilon}(t)]=0,\;
	\nabla^2\mathscr{F}[\bm{\varepsilon}(t)]<0.
\end{equation}
Therefore, a strong trap in the fitness landscape is deliberately
set~\cite{dFS13}.  The resultant Hamiltonian is obtained by
inserting~(\ref{eq:qutritcontrol}) into~(\ref{eq:controlHamiltonian}).
The external field, or control parameters, are adjusted to
realize~(\ref{eq:qtarget}) according to expression~(\ref{eq:U}).  In
this way, we can realize the qutrit phase gate by quantum control
using the control and drift Hamiltonians~(\ref{eq:qutritcontrol}).

While this problem is deliberately contrived to illustrate the
existence of traps, the drift Hamiltonian~(\ref{eq:qutritcontrol}),
for a fixed value of~$T$, effectively describes the free evolution of
a three-level system such as encountered in a spin-1 system or in a
single atom with three pertinent electronic levels.  The diagonal
terms correspond to the electronic energy levels of the atom.

The first and second energy levels are non degenerate in the absence
of the driving field, but the states approach degeneracy in the limit
of long time~$T$.  A system whose energy levels depend on the control
time~$T$ does not appear in nature but is a legitimate system for
mathematically exploring the limitations of greedy algorithms and
power of evolutionary algorithms.  Therefore, we employ this model in
the qutrit case to compare these different optimization strategies.

The control Hamiltonian~(\ref{eq:qutritcontrol}) can represent the
interaction of an atom with a driving field.  The diagonal terms of
the control Hamiltonian represent level shifts due to the effect of
the field.  The off-diagonal terms are the Rabi frequencies between
the corresponding pairs of levels, in this case between the first and
second levels and between the second and third levels.  Here we have
scaled the Rabi frequencies to unity.

\subsection{\textsc{CNot} gate}
\label{subsec:CNot}
The second example concerns the two-qubit~\textsc{CNot} gate~\cite{BBC+95}.
Inspired by the one-dimensional linear Ising-$ZZ$ model~\cite{MSG+11},
the drift and control Hamiltonians are
\begin{equation}
\label{eq:CNOTdriftcontrol}
	\hat{H}^\text{dr}
		=\frac{J}{2}Z\otimes Z,\;			
	 \bm{\hat{H}}^\text{c}
	 	=\begin{pmatrix}
			X\otimes\openone\\
			\openone\otimes X\\
			Y\otimes\openone\\
			\openone\otimes Y
			\end{pmatrix},
\end{equation}
respectively, for
\begin{equation}
	X=\frac{1}{2}\begin{pmatrix}0&1\\1&0\end{pmatrix},
	Y=\frac{1}{2}\begin{pmatrix}0&-\text{i}\\ \text{i}&0\end{pmatrix},
	Z=\frac{1}{2}\begin{pmatrix}1&0\\0&-1\end{pmatrix}
\end{equation}
the non identity Pauli matrices and $\openone$ the $2\times2$ identity
matrix.  We normalize time by setting $J\equiv 1$.  The time-dependent
four-dimensional control vector~$\bm{\varepsilon}(t)$ in
Eq.~(\ref{eq:controlHamiltonian}) is optimized so that the resultant
evolution~(\ref{eq:U}) approximates \textsc{CNot} with high~$\mathscr
F$.

Physically, the Ising-$ZZ$ model corresponds to a
one-dimensional spin chain, which was originally studied in the context of
  explaining ferromagnetism.
The weak interaction is described by a
  tensor product of Pauli~$Z$ operators for nearest neighbors.  The
  spin chain interacts with an external field, for example, a magnetic
  field, and this interaction involves only single non-identity Pauli
  operators as seen in the control
  Hamiltonian~(\ref{eq:CNOTdriftcontrol}).

\section{Results}
\label{sec:Results}

  In this section we first discuss the performance of the
  quasi-Newton method on two specific problems where there is enough time for
  unitary operation or a sufficient number of control parameters. 
  Then we numerically show that reducing the time~$T$ and control parameters~$K$ transforms the problem into a hard optimization problem and
  results in runs of the greedy algorithms getting trapped.  In the
  next part of this section we evaluate the performance of
  evolutionary algorithms when the time is shortened and~$K$ is
  reduced and compare the performance of different algorithms in terms
  of their median and best plots and finally tabulate the resultant data.

\begin{figure}
	\includegraphics[width=0.49\columnwidth]{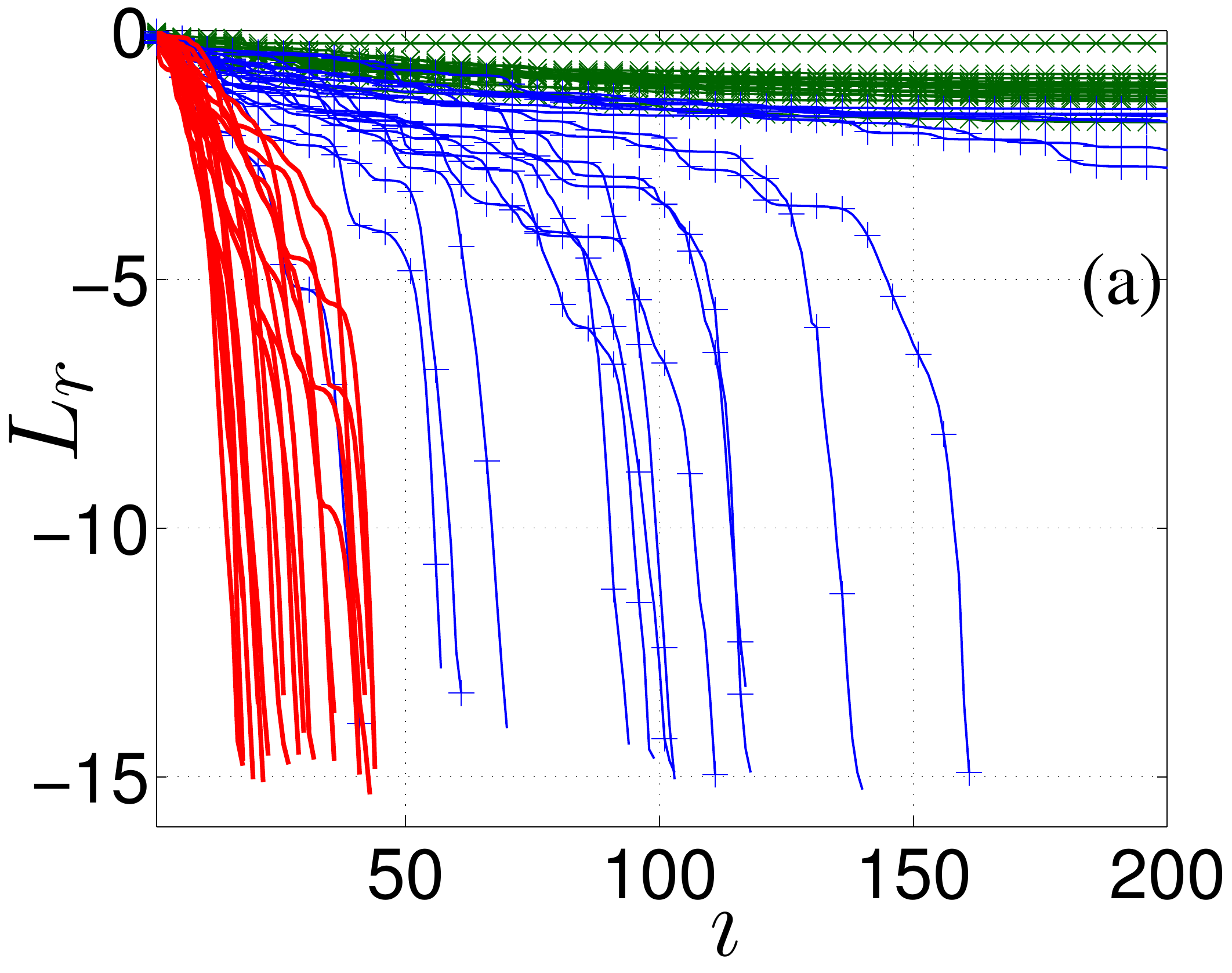}
	\includegraphics[width=0.49\columnwidth]{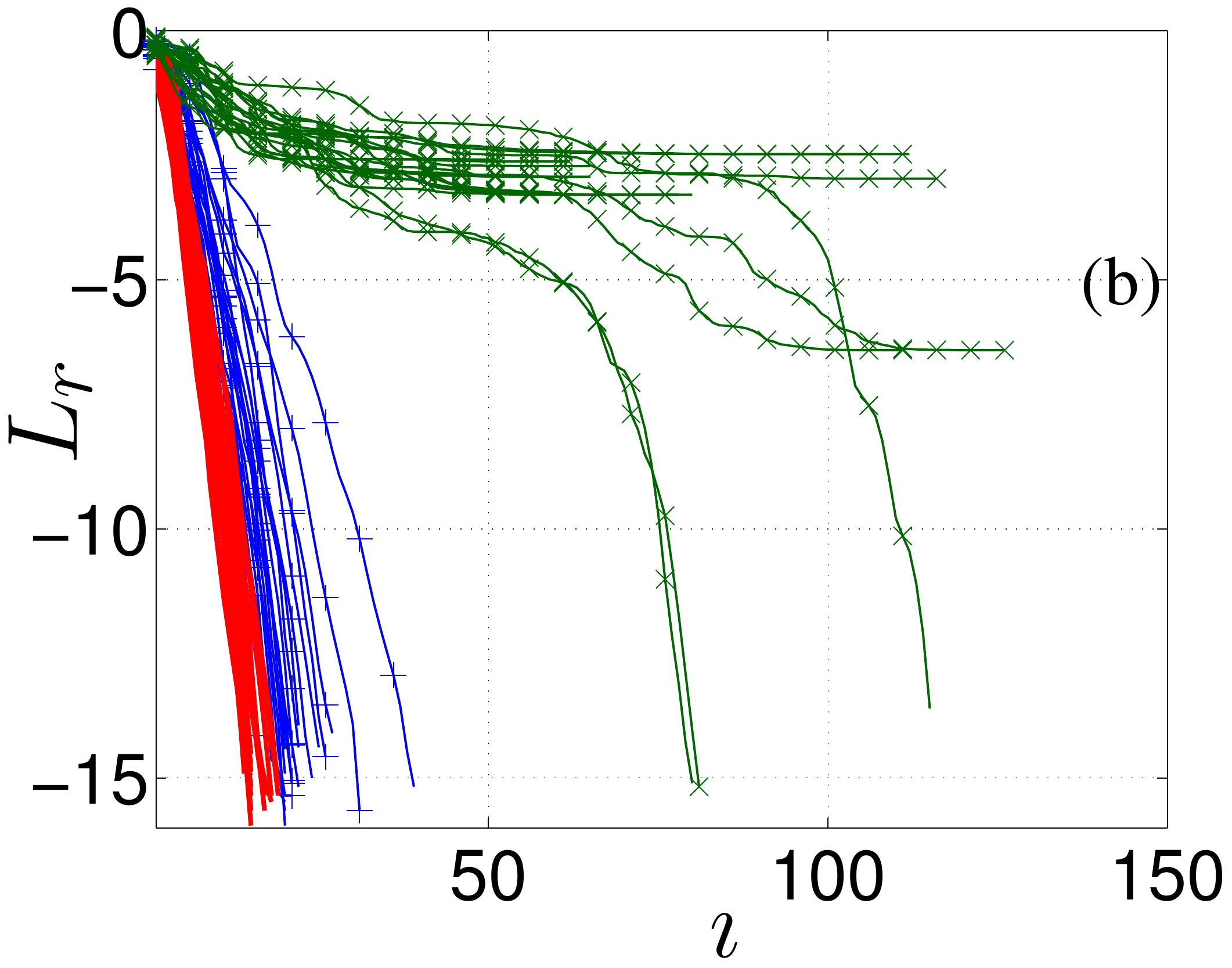}
\caption{
	(Color online)
	Logarithmic infidelity~$L$ vs iteration number~$\imath$ for
	(a)~the qutrit gate and
	(b) the \textsc{CNot} gate
	using the quasi-Newton method with 
	(a) $T=10\pi$ (red solid lines),
	$T=4\pi$ (blue lines with ``$+$'' markers), and
	$T=3\pi$ (green lines with ``$\times$'' markers) such that $K=50$ in all cases, and with
	(b)~$T=30$ and $K=30$ (red solid lines),
	$T=10$ and $K=10$ (blue lines with ``$+$' markers), and
	$T=4$ and $K=4$ (green lines with `'$\times$'' markers).
	}
\label{fig:greedy}
\end{figure}

  We choose to begin with a plot of the fast-convergent quasi-Newton
  method because this approach should be the preferred choice for when it succeeds to
  deliver a satisfactory results.
Figure~\ref{fig:greedy} depicts
the logarithmic infidelity~$\{L_r\}$ as a function of~$\imath$ for the
qutrit phase gate and for the \textsc{CNot} gate using the
quasi-Newton method.

\begin{figure}
	\includegraphics[width=0.47\columnwidth]{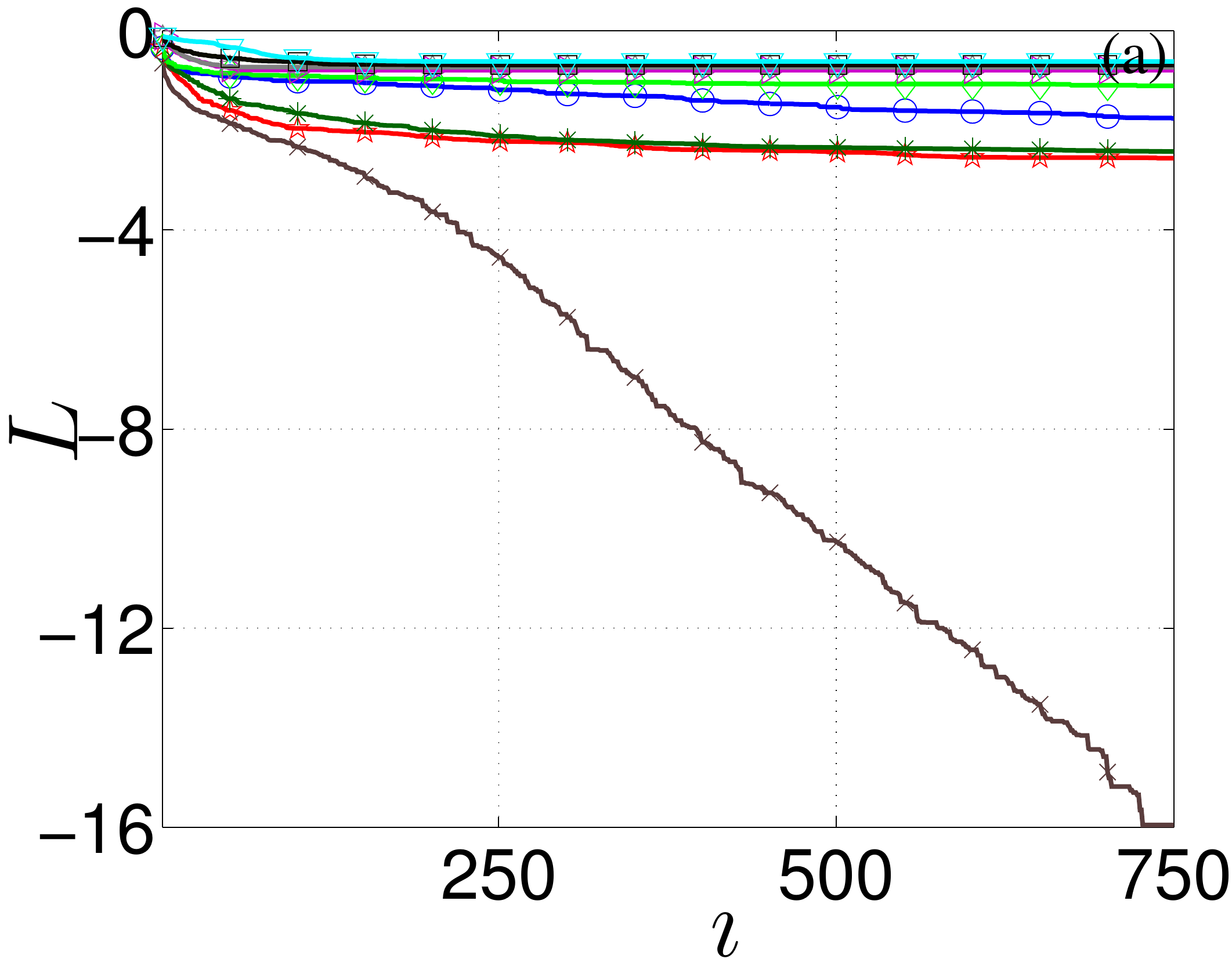}
	\includegraphics[width=0.47\columnwidth]{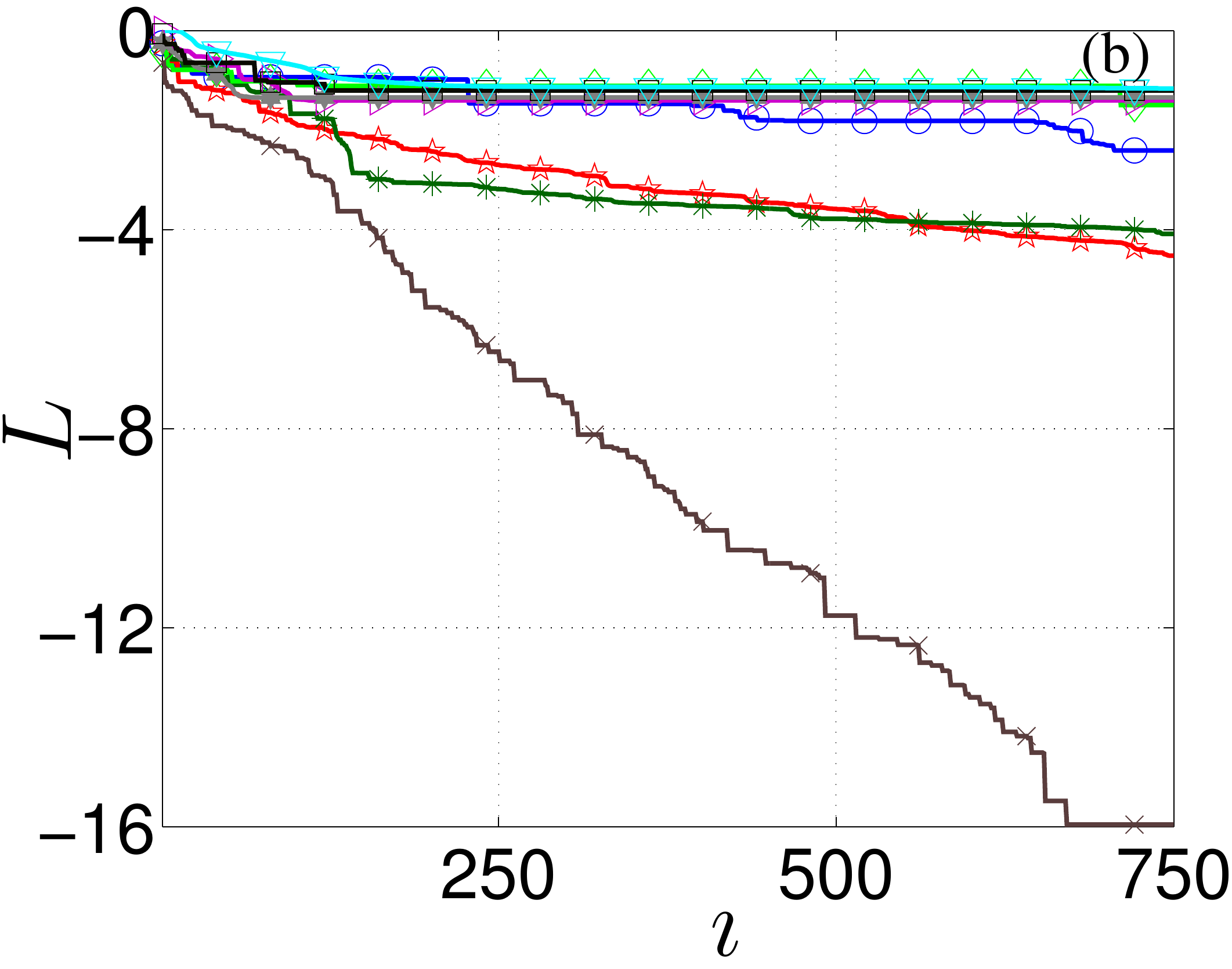}\\
	\includegraphics[width=0.47\columnwidth]{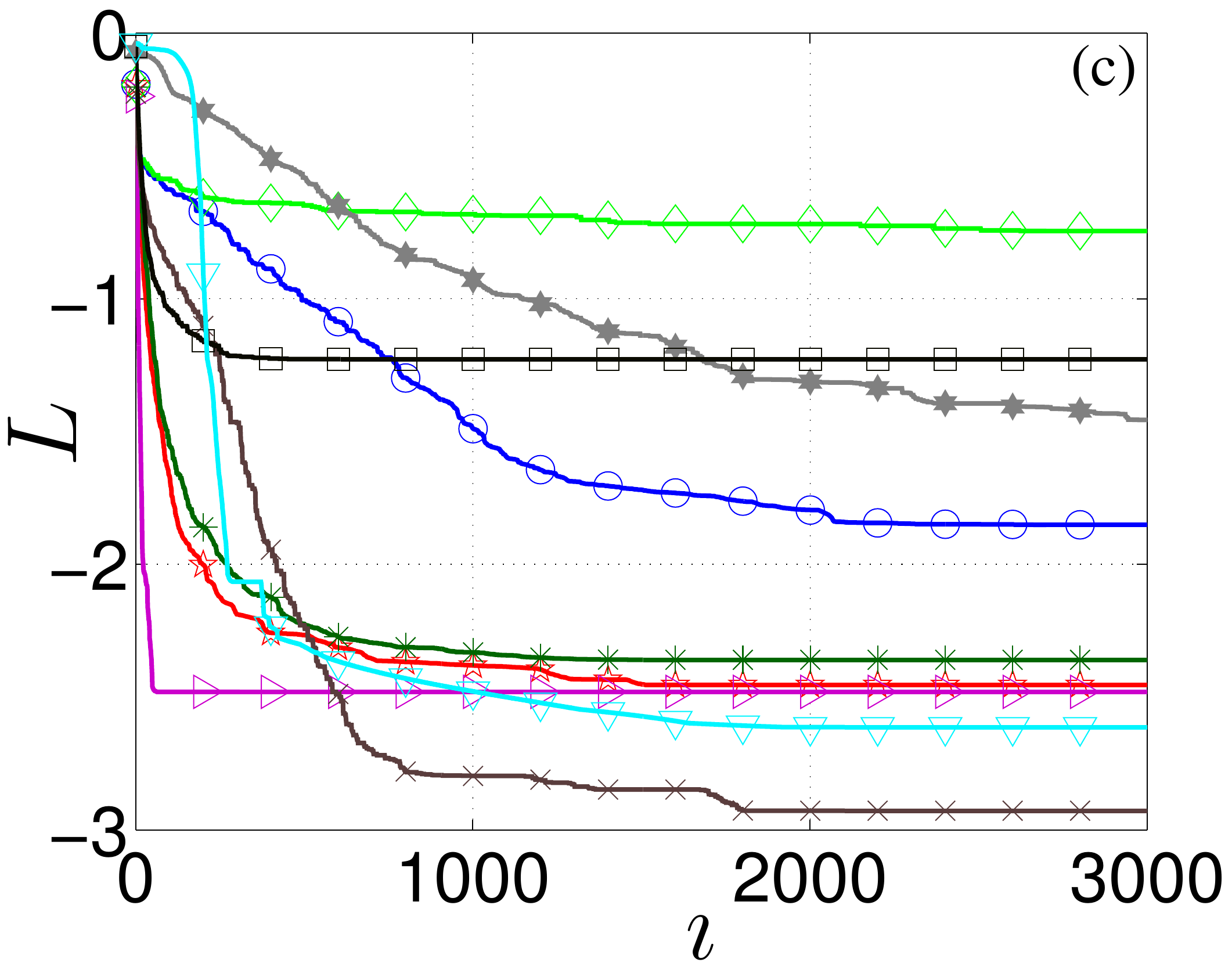}
	\includegraphics[width=0.47\columnwidth]{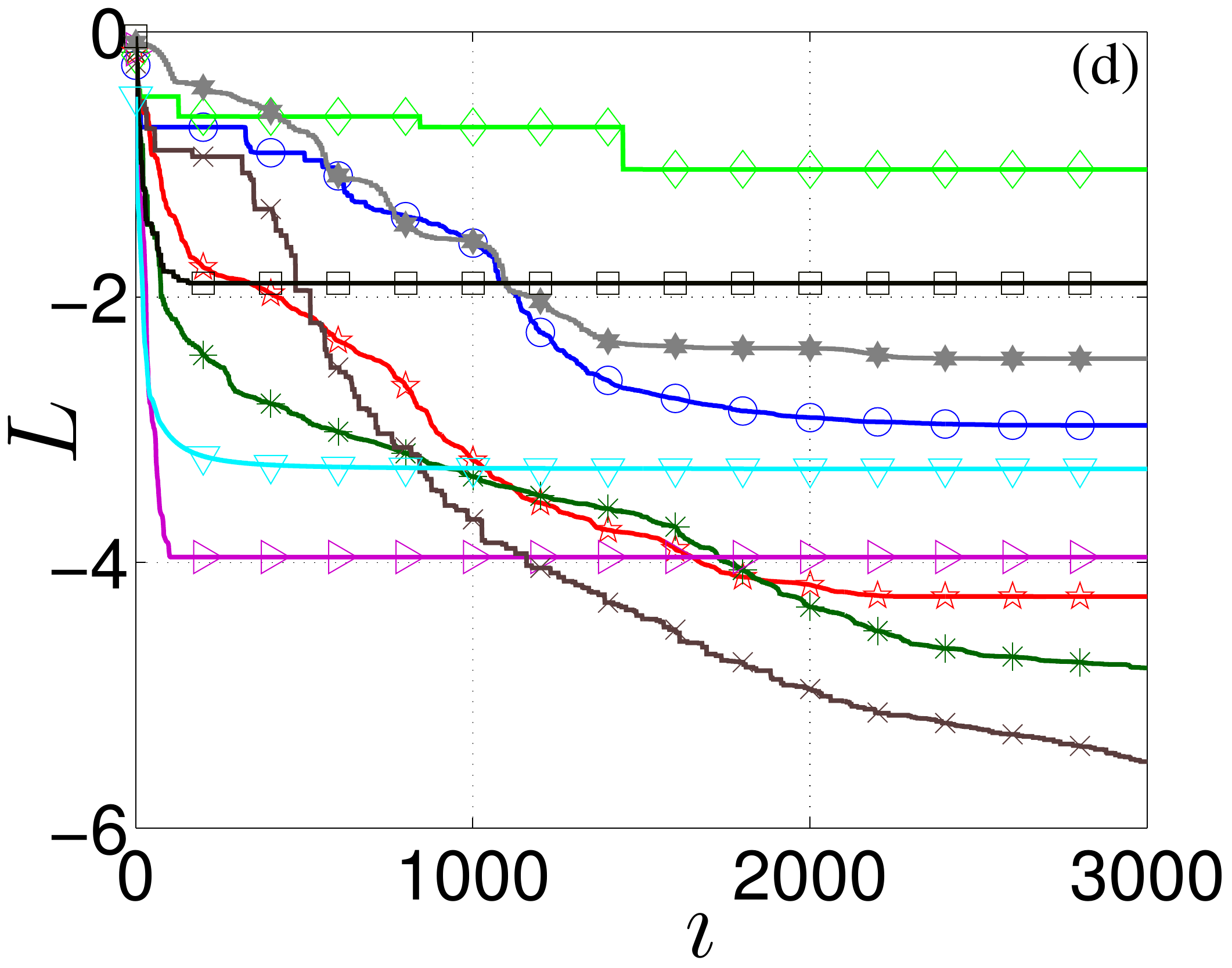}
\caption{(Color online)
        Logarithmic infidelity~$L$ vs iteration number~$\imath$ for ($\Box$)~GA, ($\times$)~DE,
	($\circ$)~Common PSO, (\ding{73})~PSO1, ($\ast$)~PSO2, ($\Diamond$)~PSO3, ($\rhd$)~quasi-Newton, (\ding{86})~simplex, and ($\triangledown$)~Krotov
	with $R=80$ ($R=40$) for greedy (evolutionary) algorithms.  Median-run performance is
	depicted in~(a) and~(c); best~run performance is depicted in~(b)
	and~(d).  
	The qutrit phase gate is the target ($T=2.5\pi$ and
	$K=10$) in~(a) and~(b); \textsc{CNot} is the target
	($T=3.2$ and $K=4$) in~(c) and~(d).
	}
\label{fig:comparison}
\end{figure}

In Fig.~\ref{fig:comparison}, we compare the greedy simplex,
Krotov, and quasi-Newton methods against~GA, DE, Common PSO, PSO1,
PSO2, and PSO3 algorithms.
Specifically, we depict best-run performance
and median-run performance in terms of final~$L$.
These plots are indicative only.  Careful comparisons are summarized in
Table~\ref{table:qutrit} for the qutrit case and in
Table~\ref{table:CNot}
\begin{table}[t]
\centering
\resizebox{\columnwidth}{!}{
\begin{tabular}{l|ccccccccc}
	$L$ & GA& DE  & PSO & PSO1  & PSO2   & PSO3  &Newton &simplex&Krotov \\\hline
	Median  &-0.6 & -15.9 & -1.7 & -2.5 & -2.4 & -1.1 & -0.7&-0.7&-0.6\\
	Best case&-1.2&  -15.9 & -2.4 & -4.4 & -4.1 &-1.5 &-1.4&-1.3&-1.16\\ 
	Worst case&-0.4& -2.2 & -1.3 & -1.6 & -1.4 & -0.9 & -0.4&-0.4&-0.4\\ 
	$\wp^\text{t}$&0& 72.5 &     0 & 12.5 &  7.5 & 0& 0& 0&0\\
\end{tabular}
}
\caption{
	Median, best case, worst case, and $\wp^\text{t}$ for logarithmic infidelity~$L$ for the qutrit phase gate with $T=2.5\pi$, $K=10$,
	and $R=80$ ($R=40$) for greedy (evolutionary) algorithms.
	}
\label{table:qutrit}
\end{table}
\begin{table}
\centering
\resizebox{\columnwidth}{!}{
\begin{tabular}{l|ccccccccc}
	$L$ &GA& DE  & PSO & PSO1  & PSO2   & PSO3 & Newton&simplex&Krotov \\\hline
	median &-1.2   & -2.9 & -1.8 & -2.4 & -2.3 & -0.7 & -2.4&-1.45&-2.6\\
	best    &-1.8 & -5.5 & -2.9 & -4.2 & -4.7 & -1.0 & -3.9&-2.4&-3.2\\ 
	worst   &-0.7 & -2.0& -1.3 & -1.9 & -1.3 & -0.6 & -2.0&-0.8&-1.9\\ 
	$\wp^\text{t}$&0 &15.0 &0 & 2.5 & 10.0 & 0  & 0& 0&0\\
\end{tabular}
}
\caption{
	Median, best case, worst case, and $\wp^\text{t}$ for  logarithmic infidelity~$L$ for the \textsc{CNot} gate with $T=3.2$, $K=4$, and $R=80$ ($R=40$) for greedy (evolutionary) algorithms.
	}
\label{table:CNot}
\end{table}
for median, best-case, and worst-case performance as well as for the
percentage of runs~$\wp^\text{t}$ that exceed~$L^\text{t}$ over $R=40$
runs for evolutionary algorithms and over $R=80$ runs for greedy
algorithms.

\section{Discussion}
\label{sec:discussion}
This section begins with a discussion about the greedy
  algorithm used for two quantum-control examples, namely, the qutrit
  phase gate and the \textsc{CNot} gate.  As we are deliberately
  making the problem harder by reducing the time and control
  resources, we resort to evolutionary algorithms and discuss their
  performances on these specific cases. We discuss the numerical
  evidence of local traps later on in this section and compare the
  performance of evolutionary algorithms with greedy algorithms when
  local traps dominate the landscape.
In the last part we compare
  evolutionary algorithms performance and discuss why DE outperforms
  its ancestor GA.

As we explained in earlier sections, greedy algorithms converge
  faster than evolutionary algorithms and they should be the first
  choice for quantum-control optimization if they can provide a
  satisfactory results.  We show the greedy-algorithm performance in
  Fig.~\ref{fig:greedy}, where,
  in both cases (qutrit phase gate and
  \textsc{CNot} gate), most quasi-Newton runs converge rapidly within machine
precision ($L=-15.65$) to the target gate for large~$T$ and for small
time resolution $T/K$.  For small~$T$ and~$K$, a majority of runs
become trapped at low fitness (high~$L$) values.  Evidently, the
quasi-Newton method fails (green plots in Fig.~\ref{fig:greedy}) for
short-time and fine--time-resolution constraints.

Our results show that greedy algorithms perform poorly for the
  highly constrained-$T$, low-$K$ problems as do PSO and
  GA. 
  Figure~\ref{fig:comparison} compares the performance of different
  algorithms for two cases by providing the median and best plots for
  each algorithm. For the qutrit phase gate,
  optimization performance is shown in Figs.~\ref{fig:comparison}(a) and~\ref{fig:comparison}(b).
  Evidently, all
  quasi-Newton, Krotov, and simplex runs become trapped at very low fidelity. 
  On the other hand, DE
  and PSO1 and PSO2 are the only algorithms that successfully achieve the infidelity target of
  $L=-4$ 72.5\%, 12.5\% and 7.5\% of the time, respectively (c f., 
  Table~\ref{table:qutrit}), and DE achieves the best performance in
  terms of the best and median infidelity among all algorithms.
  \\ \indent In the qutrit-phase-gate example we are searching the
  landscape around a critical point $\varepsilon(t)=0$ by sampling
  each trial $\varepsilon(t)$ randomly from $[-1,1]$
  and evolving them toward their optimal values.
  As there are other studies that numerically provide
  the proof of local traps in the quantum control
  landscape~\cite{PT12IJC,dFS13,PT12}, here our results show many
  local traps in the landscape as many runs from greedy algorithms get
  trapped at low fidelities. Using an efficient global-optimization
  routine like DE is necessary to avoid these local traps and to find
  a global optimal.
 
 For the \textsc{CNot} gate,
 whose performance is shown in Figs.~\ref{fig:comparison}(a) and~\ref{fig:comparison}(b),
   all runs become trapped at poor fidelities for the greedy
   algorithm case, and the~GA and various PSO algorithms are also
   poor.  In contrast the DE performance is vastly superior for the
   qutrit phase gate and significantly better for the \textsc{CNot}
   gate under the extreme conditions of $T=3.2$ and $K=4$.  Naturally,
   the greedy and PSO algorithms can be improved by increasing~$K$,
   and this strategy is common in the quantum-control literature, but
   our aim is to constrain~$T$ and limit the number of control
   parameters~$K$, and DE is the superior tool for doing so in that it
   works when the greedy and GA algorithms fail.  \\ \indent In
   the~\textsc{CNot} case, DE succeeds in providing a satisfactory result
   (see Table~\ref{table:CNot}) 15\% of the time whereas this success rate
   is 2.5\% and 10\% for the PSO1 and PSO2 cases and zero for other cases.
   Therefore, 
   this result shows that for short~$T$ and small~$K$ there are many
   local traps in the landscape causing the greedy algorithms to fail.

In all evolutionary algorithms discussed here,
DE always
  outperforms its algorithmic ancestor GA. One might ask why DE performs better
  than GA~\cite{PBC+09} on these two specific quantum-control problems
  and the answer lies in the mechanism of generating
the new population from the old population. In GA, parents are
selected based on probabilities that lead to individuals with better
fitness.  The crossover operation combines partial parts of two
parents to generate a new offspring.  As the new offspring comes from
a combination of two parents, in this sense GA explores the optimal
solution around some good solution candidates. GA must perform the
mutation operation on the individual with a low mutation probability
constant; otherwise, it turns into a searching algorithm and becomes
inefficient. This low mutation probability limits the GA's ability in
searching the whole domain of landscape and thus might cause GA to
fail with locating the global optimum.

Unlike GA, which converts candidate solutions into a binary format, DE
constructs candidate solutions that are represented by real numbers.
The crossover operation in DE generates offspring individuals from the
entire set of populations so that newly generated offspring are always
different from parent individuals. The higher mutation probability in
DE, compared to GA, enables DE to explore the search space more
efficiently while reducing the chance of getting trapped in local
minima hence outperforms GA in term of the quality of the results (see
Fig.~\ref{fig:comparison})

  Optimization strategies can be compared in various ways.  The
  most important criterion is whether the optimization approach
  delivers a satisfactory result.  A secondary consideration is the rate of
  convergence, which is relevant to the run time.  Of course, use of
  computational space is another consideration.  In our case we are
  most concerned with the primary consideration of whether the
  optimization works, as determined by whether the threshold
  infidelity reaches $L=-4$.
  
  As shown in~Fig.~\ref{fig:comparison},
  the quasi-Newton method converges faster than all other approaches but fails to achieve $L=-4$.
  Our message is that
  the most efficient, fastest optimization strategy should be used as
  long as it delivers a satisfactory result.
  If the fastest routine failed,
then our analysis
  of two tightly time-constrained control problems
is that DE is an excellent
  alternative that appears to deliver a satisfactory result
  even when the other approaches fail.
  
The fast convergence of quasi-Newton runs
     in~Fig.~\ref{fig:comparison} raises the tantalizing possibility of whether increasing the
     number of quasi-Newton runs would result in a small but non zero
     success probability~$\wp^\text{t}$.
 A fast algorithm like quasi-Newton with a low probability of success could make it superior 
 to the slow DE approach with a high success probability.
 To test this hypothesis,
we did 500 repetitions of 100 quasi-Newton iterations
applied to the (\textsc{CNot}) gate control problem.
 We chose 100 iterations of quasi-Newton runs as the average ``wall time''
 (the true run time on the given computer)
for 100 iterations approximates the average wall time for 40 iterations of the DE method.
Our numerical study showed that the quasi-Newton runs never reached $L\leq -4$.
In principle, the quasi-Newton method would work with a sufficient number of trials
simply because the global search would be achieved by a huge number of local searches,
but replacing a good global search by extremely many local searches is not feasible in practice.

Finally we emphasize that, when greedy algorithms work, the quantum
control strategy should be to employ current practice and use the best
available greedy algorithm.  When greedy algorithms fail, though,
evolutionary algorithms could work and DE is the
best among these according to our investigation.  This is
particularly relevant when exploring quantum speed limits numerically.
In view of our results, quantum speed limits found using greedy
algorithms reflect the limitations of these algorithms rather than
intrinsic speeds limits for quantum control.

\section{Conclusion}
\label{seb:Conclusion}
In conclusion we have shown that evolutionary algorithms such as DE
and PSO are essential alternatives to greedy algorithms for hard
quantum control problems with strong constraints.  Greedy algorithms
are often used because fitness landscapes are assumed to be
well behaved~\cite{RHR04}, and traps presumed to be negligible if~$T$
can be long and~$K$ can be increased without paying a significant
price.  In such cases greedy algorithms work because most local optima
are globally optimal or close enough.  However, when resources are
limited, even straightforward control problems for simple systems can
become hard due to a proliferation of traps in the landscape and
non convexity, thereby causing greedy algorithms to fail.

We have considered two quantum gates relevant to quantum information and
used drift and control Hamiltonians that illustrate our point.  These
examples show that DE is effective for hard quantum
control problems.  The superiority of DE over greedy
algorithms is unsurprising because the fitness landscape is no longer
well behaved for hard quantum control.  On the other hand, the
superiority of differential evolution over~GA and~PSO and its variants is due
to the greater efficacy of DE for optimization over higher-dimensional
search spaces, which is the case for hard quantum control.

\begin{acknowledgments}
E.~Z. acknowledges a Murray Fraser Memorial Graduate Scholarship and Eyes
High International Doctoral Scholarship and support from NSERC.  S.~S.~acknowledges support from EPSRC and EU network QUAINT.  B.~C.~S.~is
supported by NSERC, CIFAR, USARO, and AITF and acknowledges hospitality
and financial support from Macquarie University in Sydney and from the
Raman Research Institute in Bangalore, where some of this research was
performed.  S.~S.~and B.~C.~S.~acknowledge valuable discussions with
A.\ Pechen and Y. R. Sanders during the early stages of this work.
This project was initiated at a Kavli Institute for Theoretical
Physics Workshop and thus supported in part by the National Science
Foundation under Grant No.\ NSF PHY11-25915.
\end{acknowledgments}

\appendix
\label{appendix}
\section{Genetic Algorithm}
Genetic algorithms~\cite{Hol75} are well known for global optimization.
A candidate solution is first coded in a binary representation,
called a parent vector.
These parents evolve through several algorithmic steps,
namely selection, crossover, and mutation.
These steps lead to the generation of new candidates,
known as children or offspring, for subsequent generations.
These children become parents for the next generation.

Several variants of algorithmic steps exist for GA~\cite{Gol89}.
These steps evolve the fitness function towards its optimal state.
We choose the following GA variant that leads to the best performance for our problem.
\begin{enumerate}
\item \textbf{Selection:--}
This step specifies how the GA chooses the next-generation parent for subsequent breeding.
We use the roulette wheel method,
which assigns a selection probability to each individual parent vector according to
\begin{equation}
	p_i=\frac {f_i} {\sum_{i=1}^N f_i}
\end{equation}
for~$N$ the total population number and~$f_i$ the fitness level of each individual parent vector.
This probability distribution is used to select parent vectors for the crossover step.
\item  \textbf{Crossover:--}
This step, which is considered to be the heart of GA,
specifies how the two parents unite to generate the new offspring.
We use the two-point selection method to choose two random integers~$m$ and~$n$
between one and the number of variables in each parent vector.
Offspring elements are constructed from the element of the first parent vector~$P_1$,
whose indices are less than~$n$ or greater than~$m$,
and those elements of the second parent~$P_2$,
whose elements share equal indices or are between~$n$ and~$m$.
\item \textbf{Mutation:--}
The purpose of mutation is to introduce small changes in an individual selected from the population,
which leads to the creation of mutant offspring.
We mutate uniformly on each individual offspring.
The mutation algorithm thus selects vector elements that are be mutated according to a small rate of 0.001.
Then those selected elements are replaced by a random number selected uniformly
from the set of all elements of the corresponding offspring vector.
\end{enumerate}
For all problem instances,
we set $N=70$.
This choice of population number ensures the same computational time for GA as for other evolutionary algorithms here, namely, DE and PSO. 

\section{Particle Swarm Optimization}
PSO optimizes by enabling exploration of the fitness landscape using a swarm of particles
with position-velocity pairs
$\{(x_n,v_n)\}$.
These pairs are updated in each iteration of the algorithm
based on the rules
\begin{subequations}
\begin{align}
 	v_{n+1} &= \chi (w v_n + c_1 r_1 (x_{n,*}-x_n)+c_2 r_2 (x_{*}-x_n)),\\
 	x_{n+1} &= x_n + v_n,
\end{align}
\end{subequations}
where ~$x_{n,*}$ is the~$n^\text{th}$ particle's previous personal best and $x_*$ the
global best position so far. 
We employ~$\chi$ as a constriction factor,
and~$w$ is an inertial weight.
The coefficients~$c_1$ and~$c_2$ are deterministic weights, and
$r_1$ and~$r_2$ are uniformly distributed random numbers in~$[-1,1]$.

For the common PSO algorithm,
the inertial weight decreases linearly starting from
$w_{\max}=0.9$ to $w_{\min}=0.4$ over $N$ iterations according to
$w_n=w_{\max}-(n-1)(w_{\max}-w_{\min})/N$ and the standard parameters
are $c_1=c_2=2$ and $\chi=1$.  Clerc's and Trelea's variants use
constant inertial weights and different parameter values.  Clerc uses
$w=1$ (PSO1), whereas Trelea uses $w=0.6$ (PSO2) and $c_1=c_2=1.7$ (PSO3)
and $w=0.729$ and $c_1=c_2=1.492$ (variant 2).

\section{Differential Evolution}
Individuals in DE are represented by a $D$-dimensional vector $(X_i)$,
$i \in$ $ \lbrace 1,\ldots, N_\text{P} \rbrace$, where $D$ is the number
of control parameters and $N_\text{P}$ is the population size. The classical DE 
algorithm can be summarized as follows~\cite{SP97}.
\begin{enumerate}
\item \textbf{Mutation:--}
The update step is
 \begin{equation}
 \label{eq:mu}
 V_i=X_{i_1}+\mu\left(X_{i_2}-X_{i_3}\right),
 \end{equation}
with~$i$, $i_1$, $i_2$, $i_3$ $\in$ $[1,N_P]$ being integers and mutually
different.
Here~$\mu$ is the mutation factor controlling the
differential variation $d_i:=X_{i_2}-X_{i_3}$.
\item \textbf{Crossover:--} 
\begin{equation}
\label{eq:cr}
	C_i(j)= \begin{cases}
	V_i(j) &\text{if $C_j(0,1)<\xi$},\\
	X_i(j) &\text{otherwise},
\end{cases}
\end{equation}
with
	$C_j(0,1)$ representing the uniform random between 0 and 1,
	  and~$\xi \in (0,1)$ is the crossover rate.
\item \textbf{Selection:--}
The final step is the assignment
\begin{equation}
\label{eq:sl}
	X'_i= \begin{cases}
	C_i &\text{if $f(C_i)<f(X_i)$},\\
	X_i &\text{otherwise},
\end{cases}
\end{equation}
with~$X'_i$ the offspring of $X_i$ for the next
generation and $f(X_i)$ the objective function,
which, in our case, is the measured fidelity.
\end{enumerate}
For all instances we choose $\mu=0.5$,
$\xi=0.9$, and $N_\text{P}=15K$, 
with~$K$ being the number of
control parameters, for all problems.

%
%
\bibliographystyle{apsrev}
\bibliography{qcontrol}

\end{document}